# Theory-Software Translation: Research Challenges and Future Directions

Caroline Jay, Robert Haines, Daniel S. Katz, Jeffrey Carver, James C. Phillips, Anshu Dubey, Sandra Gesing, Matthew Turk, Hui Wan, Hubertus van Dam, James Howison, Vitali Morozov, Steven R. Brandt

**Abstract**
The Theory-Software Translation Workshop, held in New Orleans in February 2019, explored in depth the process of both instantiating theory in software – for example, implementing a mathematical model in code as part of a simulation – and using the outputs of software – such as the behavior of a simulation – to advance knowledge. As computation within research is now ubiquitous, the workshop provided a timely opportunity to reflect on the particular challenges of *research software engineering* – the process of developing and maintaining software for scientific discovery. In addition to the general challenges common to all software development projects, research software additionally must represent, manipulate, and provide data for complex theoretical constructs. Ensuring this process is robust is essential to maintaining the integrity of the science resulting from it, and the workshop highlighted a number of areas where the current approach to research software engineering would benefit from an evidence base that could be used to inform best practice.

The workshop brought together expert research software engineers and academics to discuss the challenges of Theory-Software Translation over a two-day period. This report provides an overview of the workshop activities, and a synthesises of the discussion that was recorded. The body of the report presents a thematic analysis of the challenges of Theory-Software Translation as identified by workshop participants, summarises these into a set of research areas, and provides recommendations for the future direction of this work.

# Table of Contents

# 1. Introduction

In 2017, the Software Sustainability Institute-funded Code/Theory Workshop[1] took place in the UK. The aim of the workshop was to explore the challenges involved in writing software for research. The topics emerging from the discussion were broad, ranging from technical issues to organisational culture.

The results of this workshop made it clear that an exploration of the unique technical challenges of research software engineering was required. The [Theory-Software Translation Workshop](#), held in New Orleans in February 2019, examined this issue with a particular focus on High Performance Computing (HPC). The term 'Theory-Software Translation' was used to cover in broad terms the processes involved in implementing, and understanding the results of computation within research. As an example of the challenges involved, consider interpreting the results of a simulation: it is necessary to distinguish between underpinning theory (knowledge), mathematical model and numerics (applied mathematics) and implementation (code). Moving between these representations requires abstraction, and therefore entails loss and compromise. Assumptions made when translating a model to numerics and then to code may have a profound effect on the simulation behaviour, but may be poorly understood by a domain scientist using the results, due to a lack of expertise, and the complexity of the underlying codebase.

The workshop explored Theory-Software Translation in general terms, but with a particular focus on understanding how empirical research could contribute to improving it. Computation is now ubiquitous in science, but its implementation has evolved largely in an ad hoc and incremental fashion. By examining the process critically and in detail, the workshop aimed to determine the perceived threats to scientific validity posed by current practices, and also to identify the evidence and/or activities that could lead to a step-change in the validity and robustness of the knowledge produced via computational science.

## 1.a. Participants

The following experts within the fields of research software engineering, and the academic study thereof, attended the workshop by invitation:

Roscoe A. Bartlett, Sandia National Laboratories, USA
Steven R. Brandt, Louisiana State University, USA
Jeffrey C. Carver, University of Alabama, USA
Thomas Cheatham, III, University of Utah, USA

---

[1] Jay C, Haines R, Vigo M, Matentzoglu N, Stevens R, Boyle J, Davies A, Del Vescovo C, Gruel N, Le Blanc A, Mawdsley D, Mellor D, Mikroyannidi E, Rollins R, Rowley A, Vega J (2017) *Identifying the challenges of code/theory translation: report from the Code/Theory 2017 workshop.* Research Ideas and Outcomes 3: e13236.
https://doi.org/10.3897/rio.3.e13236

Anshu Dubey, Argonne National Laboratory, USA
Sandra Gesing, University of Notre Dame, USA
Rinku Gupta, Argonne National Laboratory, USA
Robert Haines, University of Manchester, UK
James Howison, University of Texas at Austin, USA
Caroline Jay, University of Manchester, UK
Hans Johansen, Lawrence Berkeley National Laboratory, USA
Daniel S. Katz, University of Illinois at Urbana-Champaign, USA
Dmitry Liakh, Oak Ridge National Laboratory, USA
Vitali Morozov, Argonne National Laboratory, USA
Brian O'Shea, Michigan State University, USA
James C. Phillips, University of Illinois, USA
Katherine Riley, Argonne National Laboratory, USA
Matthew Turk, University of Illinois at Urbana-Champaign, USA
Hubertus van Dam, Brookhaven National Laboratory, USA
Hui Wan, Pacific Northwest National Laboratory, USA

## 1.b. Workshop format

The workshop started with an introduction to the topic from the organisers, followed by a short talk on the 'Four Facings' from Tom Cheatham, which examined the different perspectives that should be considered when evaluating research software. Participants then gave lightning talks explaining their background and interest in the topic. Slides for the lightning talks (for those speakers who used slides) are available at: https://se4science.org/workshops/tst-us/talks/.

The main part of the workshop consisted of a series of breakouts. The first, where participants were pre-allocated to groups to ensure each group had people with a mix of backgrounds, focused on defining the overall challenges of theory-software translation. Following a feedback session, and noting the themes that were starting to emerge, the organisers divided the next set of breakouts into groups considering *training and culture*, *software design*, *software stack and tools* and *miscellaneous* (to catch any issues falling outside the first three). Participants self-selected to join one of these groups for one session, and then moved to a different group for the next session. In each case, participants were asked to discuss the topic, list challenges, identify current successes, and indicate how we could make progress.

A final plenary session considered the prospects for Theory-Software Translation as a research area, and considered next steps. The full workshop agenda is available at: https://se4science.org/workshops/tst-us/agenda.

# 2. Outputs of discussions

During the breakout sessions, groups were asked to keep a record of their conversation, transcribing as much of the discussion as possible, and then summarising key points at the top of the document. There were three breakout sessions, each with four groups, resulting in 12 discussion documents. As common topics arose across groups and sessions, the breakout notes were treated as a single corpus during analysis. Two of the authors (Jay and Haines) performed a thematic analysis, with Jay categorising the full set of discussion notes into the first set of themes, Haines reviewing these and cross-checking with the discussion notes, and both iteratively refining the final set.

Three overarching themes — *Design*, *Infrastructure* and *Culture* — emerged during the analysis. We discuss these in more detail below, providing a comprehensive summary of the discussion, and describing the key areas of research identified within each. Many of the research challenges are phrased as open questions.

## 2.a. Design

Participants considered *Design* in terms of software design, research design, and the way in which the two interact.

### Should software be readable from theory?

There was considerable discussion about the extent to which it is possible to translate theory to software whilst maintaining the essence, understanding, and readability of the underlying theory. Software is highly complex, and can unintentionally obfuscate the theory it contains, particularly when it is optimized for good performance. This led to a number of questions: Is there a particular design process that should be used for embedding theory within software such that it is readable? To what extent is it necessary for someone reading the software to understand the underpinning theory? When software is composed of many components having their own theoretical underpinnings, what does it mean for the overall theory underlying the whole? Is it possible to measure how well software represents theory? Ought theory be recoverable/reconstructable from software? How could 'theory recoverability' be measured?

### Can/should we separate concerns?

In an era of growing complexity in models and questions, translating mathematical theories to software in a reliable way is becoming increasingly difficult. One perspective on the process is that of moving from 'science' to 'equations to be solved' to 'computational algorithms/numerical analysis' to 'computer science/software engineering'. (See Babuska and Oden[2] for a formal description of this process and these domains.) Each of these is a discipline in its own right, and each is complex. There was a view that it is not realistic for every scientist to understand all of these, and thus an informed 'separation of concerns' is crucial. Without this, the resultant software generally does not adequately separate the scientific questions asked, the equations to be solved, the numerical methods used to solve these equations, and the software infrastructure supporting these methods.

---

[2] Ivo Babuska, J. Tinsley Oden, "Verification and validation in computational engineering and science: basic concepts" Comput. Methods Appl. Mech. Engrg. 19:3(2004) 4057-4066

An alternative view was that concerns cannot always be separated within computational research, from both a theoretical and a practical perspective. At present, a paper and a code are separate things, but the boundaries are blurring. Jupyter notebooks are an example of documentation executed with code, but this approach is unlikely to be sufficient or scalable on its own. If the boundaries between paper and code increasingly overlap, then it becomes difficult to see where the theory ends and the software begins. Simply marking up theory in code by adding documentation increases the maintenance cost of the code and risks the two becoming out of sync. A code marked up with the wrong theory is worse than useless, even dangerous, so any such system would need to be able to verify that the code and theory were consistent.

## How can we better link domain science and computer science?

There appears to be a disconnect between computer science research and its deployment in scientific discovery; improving the linkage could lead to better science. There are many areas that require computer science research: new languages; more flexible operators; code generation; code transformation; test generation. Theory-Software Translation was recognised as having the potential to expose and contribute to these challenges.

## What is the influence of software in driving experimentation?

A researcher can use an application to verify a hypothesis through software as a first step, and then go back and modify the theory – and the software – to do more discovery and push the boundaries of the theory.

There is the potential for a broader impact of application of the theory (or theory component) into other domains, once the capability is developed for a common subtype of problem. An example of theory emerging from software is the equation of state "discovery" from brute-force simulations. How should it be merged back into theory?

In the future, code generators may offer a route to translating the theory to software. This approach could do a better job of preserving information during implementation and lead to a higher order transformation, due to higher order input. It may allow for timely cross-code validation, where different theory comparisons are made, as it is less human-resource-intensive. This may also be a way to reduce human error (for example, there are chemistry codes that do code generation where it would otherwise be error prone or repetitive), although it should be noted that code generators, being software themselves, may also introduce bugs. Recording the provenance of the code is important in understanding how theory is ultimately arrived at through software outputs. Does using a code generator obfuscate that provenance, or make it clearer?

## How can we evaluate the design process?

When trying to understand the results of a simulation, it is necessary to distinguish between theory, model, numerics (applied mathematics) and implementation (code). Forming problems at a higher level declaratively could help with this, and disciplines that make theory to software translation more reliable might be able to articulate a higher level of confidence in a simulation by focusing on testing simulations and model intercomparison. But model intercomparison is challenging, and it can be difficult to identify the reasons for and sources of similarities and differences in model outputs.

Understanding how to make comparisons between models would benefit from an empirical approach. This raises a number of additional questions: In order to be able to compare models, do we need a process for working out the resolution, or complexity, of the model? How can we avoid being too simplistic, whilst being mindful of resources? How can we avoid overfitting? Could we verify models by varying the resolution and seeing whether we get the same results? How do we deal with differences between models and real world data?

If we can gain an understanding of how models and codes can be compared, this may open up further opportunities for adaptation and reuse. Could we adapt existing codes to new paradigms? Domain Specific Languages (DSLs) are generally community specific at present. Could we make progress through merging or integrating them, at least where we can be reasonably certain that the models that they are representing are comparable? There is an explosion of tools and services across all domains. How can we tell if they are reliable? Would being able to compare them across domains help with verification and validation of these tools?

Should there be more emphasis on a clean, extensible software design from the beginning, with more extensive use of design formalization tools, like UML? Software can be used for exploration - what is the engineering process for this?

Sometimes, even when tools (e.g., machine learning frameworks, libraries) are available, the work needed to use them (transform inputs as needed, transform outputs as wanted) is too demanding. What are the design decisions that have led to this case of affairs?

## 2.b. Infrastructure

Theory-software translation is not solely about mapping scientific constructs to algorithms, but routed in and affected by a wider software and hardware infrastructure. This part of the discussion gave consideration to verification and validation, sustainability and portability, and how theory-software translation might be measured.

How should we verify results arrived at through computation?

There is currently no established, efficient means of achieving the verification of a software simulation, and as such this is an area that requires further attention. Where there is unexpected behavior in a simulation, software and data provenance are both crucial to knowing whether it is caused by a bug or a discovery. Where there is a defect, how can we tell where it lies? Is it in the theory, or the mathematics, or the code? Knowledge is required, not just of the code and the theory, but of the full software stack, for example the sequence of dependencies, and how the code compiles or is interpreted.

The number of potential inputs to most codes is much larger than can be tested in their entirety. A potential solution might be to express theory as a set of tests for code to pass. This led to the question of whether it would be possible to automate test generation from theory specification.

## Is sustainability and portability important?

The importance of reproducibility within research is becoming increasingly recognised. The extent to which true reproducibility is possible in computational science is not clear, due to portability problems, continually changing technology and software decay. Nevertheless, it was seen as important to strive to get as close as possible to this ideal, and also to work out practical ways of achieving something that approximates this. Having different teams trying to reproduce results, through multiple people running the same codes, could be useful in terms of verification and building knowledge.

Whilst sustaining software for reproducibility is difficult, paradoxically, software almost always lives longer than planned, as (for example) adding features to a prototype is quicker and cheaper than engineering a new and robust code from scratch. What are the implications of this for theory-software translation? What are the effects on the software's integrity, the way new theory must subsequently be implemented, and the results it produces? What are the issues caused by technical debt?

To address sustainability, one approach would be to develop ways to represent models and translate from code back into theory in a way that will survive over time.

## What are the constraints posed by platforms and architectures?

Scientific software is generally going to be utilized on multiple generations of computational architectures, and the original developers of the software typically do not (and cannot) take this into account. Changing hardware, as well as over-engineered codes, build systems, and supporting infrastructure impedes both portability and reproducibility and the ability to update or the theory it contains. Where concepts or operations do not currently translate to hardware, the view was that we should ideally aim to change the hardware, rather than restrict the theory. We should also remain mindful of that hardware, as well as software, can be an error source, as code that functions correctly on one platform may not on another, unbeknownst to the programmer.

## Measuring theory-software translation

Whilst theory is exact, code has tolerances and estimations. Recognising this was seen as an important part of understanding and improving theory-software translation. One suggestion was to frame this in terms of implementation decisions introducing uncertainty, and measuring how likely an output was to be correct. Rather than assuming, 'this output is correct,' would it be better to state, 'there is x% chance some error has been introduced along the way, according to the architecture/code size etc., and therefore we should interpret the result accordingly.'? Another way of considering this is from the perspective of verification. For example, could we develop diagnostics that verify the health of the simulation?

There is also loss when moving between representations in different domains (theory, equations, algorithms, software.) How can we measure this, and understand its effects?

## 2.c. Culture

The environment in which Theory-Software Translation takes place was recognised as a key influence on the process. Discussion relating to this topic covered training, collaboration, and a recognition of different backgrounds and skill sets.

### How can we foster a culture of collaboration?

Computational science, particularly that conducted in large projects, is necessarily interdisciplinary. The heavily domain-contextual specification of the problem and the deep technical knowledge required to implement solutions can lead to an initial communication barrier between domain scientists and computer scientists. Embedding software engineers in research teams is a good way of facilitating communication, and there was discussion about what more could be done.

One question was whether explicitly recognising the idea that software is a translation of theory might change the communication process. Could conversations across different roles be improved using this approach?

Implementing theory in code was viewed as different from implementing non-research software, especially where the requirements are concerned. A key issue was that it may not be possible to separate specification from design, a situation analogous to building an aircraft in flight. Given this, there is a lack of clarity about the best way to approach requirements engineering within research projects.

Crucially, it was viewed as important to emphasise that software engineering is a core intellectual contribution to the research, not just a service. Close interaction between an application scientist and an applied mathematician can be helpful in designing the appropriate mathematical/numerical method. This interaction was thought to illustrate the tension between the 'separation of concerns' that should be true within the software, but is not necessary between people. Separating concerns too strictly may lead to different people concentrating on their own tasks, with their own goals and motivations, neglecting the overall picture. On the other hand, focusing on a particular aspect can provide better abstractions and more performant solutions. How do we balance these two pressures?

There needs to be an improved understanding of which parts of a software tool can be treated as a black box and taken on faith, and which cannot. Without this understanding, software may be used in ways it is not designed for and so give spurious results. Software can be flexible, and because of this, be used in domains for which it was not originally intended; in this case, it should be validated within this new domain before any results are published.

## What are the external expectations of the reliability of the software?

Validation, which was discussed extensively from a technical perspective, was also considered from an administrative/organisational perspective. Software may need to be considered as a scientific instrument that needs to be validated and/or calibrated. A current example of this is that in the UK, any software that collects patient symptom data, that can be used to access medical advice, or that can be used to assist with a diagnosis, must be developed as a "medical device"[3]. Might there be a requirement to think of software as an instrument that meets formal standards in other research settings? Would this make results more reliable, or would it stifle creativity? Can we expect complete 'accuracy'? If not, should there be 'guards' or 'contracts' to detail this?

## How does the research environment affect the translation process?

There is a perception that academic researchers are under pressure to publish at all costs, diminishing the attention paid to good software engineering practices when they are perceived as slowing the research and publication process. Valuing software as a deliverable in addition to publications was viewed as an important part of improving its quality and availability. Citing software (via, e.g., the Journal of Open Source Software[4] or by more direct citations to the software[5]) is another part of this process. Considering software explicitly as an output of research is still relatively unusual, and there is still work to be done in understanding how to achieve this.

There was a view that funding bodies should be involved in discussions regarding theory-software translation. Many of the costs of development, maintenance, and evolution are hidden; they need to be articulated, and be part of an open, ongoing conversation. The cost of development is underestimated higher in the chain. A lot of time is spent porting software to new hardware, but it is difficult to obtain funding for this, with a negative impact on the quality of the resultant software.

## What is the best way to embed software engineering skills in science?

Often the people writing scientific code are graduate students or researchers who do not have a background in software engineering. Data Carpentry/Software Carpentry was viewed as a good start, but not sufficient. Instilling the necessity of thinking about theory-software translation in graduate students right from the start would help to avoid the need to continually apply patches on top of poorly-written and poorly-designed code. While this lack of training is a specific problem; software development training is a general challenge, because academic supervisors do not necessarily see the value of it, or even know about it themselves. Knowledge of training, and belief in the necessity of the training, is critical.

---

[3] Medical device stand-alone software including apps:
https://www.gov.uk/government/publications/medical-devices-software-applications-apps
[4] The Journal of Open Source Software: https://joss.theoj.org/
[5] Smith AM, Katz DS, Niemeyer KE, FORCE11 Software Citation Working Group. 2016. Software citation principles. PeerJ Computer Science 2:e86 https://doi.org/10.7717/peerj-cs.86

Training in communication is seen as essential, and it should go both ways. All members of a science team need to be proficient in cross-disciplinary communication. Being able to communicate scientific requirements to software developers is essential. Being a careful and skeptical user of simulation outputs is also essential, and this requires an understanding of the workings and limitations of the software method, such that simulation outputs are viewed through the appropriate lens.

There is potential for training to be conceptualised as a hierarchy: an awareness of theory-software translation at the basic level (abstract/strategy level); some general sense of how theory is translated into software and the process at a higher level (an understanding of process); and in-depth expertise in the implementation of theory-software translation at the developer level (with possible specialization).

### How do we ensure usability?

Preserving a balance between usability and performance can be difficult. Optimized code is often less readable, and so is harder to understand. This in turn further obfuscates the theory that the code represents and so impacts reproducibility and productivity as well. In an ideal project, mathematical concepts are expressed clearly in software components, offering reuse, support for testing, and a clear map to and from the underlying theory. Modular representation of theory is likely to be more readable and testable, but it would be interesting to investigate whether there are areas where this approach is not suitable. Is there value in an unoptimised, understandable version of a simulation, for example, being used as a reference implementation?

# 3. Theory-software translation as a new stream of research

The material gathered from the workshop revealed a number of areas where research could significantly advance both knowledge and scientific practice.

Theory-Software Translation research can contribute to the evidence-base for research software infrastructure strategy and practice at a local (individual/group), institutional (organization), national, and international level. It can also identify crosscutting challenges for computational research, and advance the techniques we use to perform such research. A better understanding of this process, and the ways in which errors can propagate through the various translations required, will lead to more robust and accurate research software.

Research software is not merely used to perform a task, but to understand a problem and advance knowledge. While current software engineering research outputs and methods are relevant to addressing these challenges, theory-software translation requires us to tackle new problems that are rooted within the scientific domain:

- The translation process moves from theory to algorithm to software (and vice-versa). Information is lost in moving from one domain to another, as the way in which ideas are represented changes. Can we quantify/explain this loss/difference, and articulate the tradeoffs resulting from translation?
- How does incorporating theory in software (e.g., a simulation) differ from standard requirements engineering? The development of software and theory happen together. While requirements changes are generally constrained for typical software, theory can change much more dramatically, resulting in not just an addition, but a fundamental change in what the software should do.
- How do we understand the results of a simulation, and translate this back to the theory underlying the simulation? How can we articulate confidence in it? Can we develop diagnostics that verify the health of the simulation? How should real world data be used in the verification process?
- How do we go from viable theory to validated, verified code in a time-efficient way? Can theory be expressed as a set of tests?
- There is a distinction between theory, model, numerics and code, and there are difficulties mapping between them. There can be errors in any, all, or the mapping process that may affect the resulting science. How can we detect and handle these errors?
- Is a true separation of concerns (theory, and its implementation in software) possible? What are the implications for how we write scientific software?
- Can we measure how well software represents theory? Should theory be recoverable from software (where software includes documentation)? Can theory today be represented solely in papers, or is it really also in the code? How can we help people to read and understand it?
- Are there languages or language features that make theory-software translation better/easier?
- The increase in model complexity and sophistication of questions that models are designed to answer makes translating them into software increasingly difficult. How do we determine the appropriate level of complexity? Is it possible to optimise for complexity reduction, as well as performance?
- How can we define functional reproducibility? Long-term reproducibility is likely to remain out of reach due to software collapse[6]. What should the process be? What is the role of testing? Is there a way of providing a reference implementation that can be used as a blueprint for the theory?
- Can we create ways of representing theory that will persist and remain usable, so as to increase software sustainability and prolong the period of reproducibility?
- Can we identify the range of practices currently used, and gather empirical data concerning their efficacy, to result in evidence-based best practice?
- How should we approach the design of code generators and DSLs?
- How should theory be communicated to developers? E.g, using an ontology, or formal methods?
- Software is unlikely to ever be 100% 'correct'. Can we develop ways to quantify uncertainty for software functioning such that appropriate probabilities can be applied to results?
- Can we extract theory from existing codes which may or may not be in use? How do we expose theory within complex, iteratively developed computational methods, such that scientists can understand its functioning?

---

[6] K. Hinsen, "Dealing With Software Collapse" in Computing in Science & Engineering, vol. 21, no. 03, pp. 104-108, 2019. doi: 10.1109/MCSE.2019.2900945

- If we consider the route from scientific experimental design to software and observational data, and then back to theory, can we understand where errors or discrepancies are most likely to occur? What form do these errors take at each stage of translation?
- What is the impact of making theory compromises and simplifying assumptions for the purpose of computability? Does how theory is discretized have an impact?
- What are the long term benefits of doing theory-software translation better? Can we increase scientific impact, productivity, software quality, software adoption, reuse?

# 4. Next steps

The reports from this workshop and the Code/Theory 2017 workshop identified a series of challenges in building and using scientific software. To address them, we need empirical research. We propose the following actions to help move towards this goal.

1. **Expand the community interested in this topic.** The US and UK are supporting nascent research communities in this area via URSSI and the SSI, and we believe that this growing interest and the importance of these challenges are sufficient to demonstrate the need to support high-quality research with the potential for real impact on computational science. An important aspect of this is the interdisciplinary nature of the research: tackling the challenges identified here will require a diverse range of skills.

2. **Develop an agenda for research.** Summarising the research challenges in a short publication (based on this report) will help with dissemination to a wider community, and catalyse further conversation about the best way forward.

3. **Engage with funders.** Theory-Software Translation can be perceived as 'meta-research', and therefore sits outside of traditional funding routes. It is important to make the case to funders that formally understanding the nature of computation within research 1) contains unique challenges (i.e., they are not covered by current software engineering research) and 2) has the potential to transform the scientific process.

# Acknowledgements

We thank NSF for support of this work under award 1551592.